\documentclass[twocolumn, reprint, amsmath,amssymb, aps, pre, floatfix]{revtex4-2}
\usepackage{graphicx}
\usepackage{amsmath,amssymb}
\usepackage{physics} 
\usepackage[range-phrase = --, print-unity-mantissa = false, exponent-product = \cdot, per-mode = single-symbol, uncertainty-mode = separate, inter-unit-product = .]{siunitx}
\usepackage[normalem]{ulem}
\usepackage{fontawesome}
\usepackage{xcolor}
\usepackage{ifthen}

\newcommand{\revise}{}
\newcommand{\revisecross}[1]{}

\renewcommand{\emph}[1]{\textsl{#1}}
\newcommand{\etal}{\textsl{et al.}}

\begin{document}

\title{Magnetic levitation in the field of a rotating dipole}
\author{Grégoire Le Lay}
\author{Sarah Layani}
\author{Adrian Daerr}
\author{Michael Berhanu}
\affiliation{Université Paris Cité, CNRS, Laboratoire Matière et Systèmes Complexes UMR\,7057, 75231 Paris cedex 13, France\\}

\author{Rémy Dolbeault}
\author{Till Person}
\author{Hugo Roussille}
\author{Nicolas Taberlet}
\affiliation{Univ Lyon, Ens de Lyon, Univ Claude Bernard, CNRS, Laboratoire de Physique, F-69342 Lyon, France\\}

\date{}

\begin{abstract}
It is well known that two permanent magnets of fixed orientation will either always repel or attract one another regardless of the distance between them. However, if one magnet is rotated at sufficient speed, a stable position at a given equilibrium distance can exist for a second free magnet.{The equilibrium is produced by magnetic forces alone, which are strong enough to maintain a levitating state under gravity.}
We show that a stable levitation can be obtained when the rotating magnet is tilted from the rotation axis, with no offset in its position. In this regime, the levitating magnet remains centered and its spinning rate remains negligible, while its magnetic moment precesses in synchronization with the driving magnet. We provide a physical explanation of the levitation through a model relying on static dipolar interactions between the two magnets and present experimental results which validate the proposed theory. 
\end{abstract}

\maketitle

\newlength{\figwidth}
\setlength{\figwidth}{86mm}

\renewcommand{\vr}{\vb{r}}
\newcommand{\vk}{\vb{k}}
\newcommand{\vi}{\vb{i}}
\newcommand{\vj}{\vb{j}}
\newcommand{\di}{\delta_i}
\renewcommand{\dj}{\delta_j}

\newcommand{\mr}{m_{\text{r}}}
\renewcommand{\ml}{m_{\text{l}}}
\newcommand{\Iperp}{I_\perp}
\newcommand{\Ipar}{I_{\parallel}}
\newcommand{\spin}{\omega_\text{s}}
\newcommand{\fspin}{f_\text{s}}
\newcommand{\req}{{r_{\text{eq}}}}
\newcommand{\phieq}{{\polarAngle_{\text{eq}}}}
\newcommand{\fosc}{f_\text{osc}}
\newcommand{\csteM}{G}

\newcommand{\polarAngle}{\theta}

\newcommand{\rk}[1]{
{\colorbox{orange}{\textcolor{white}{#1}}}
\marginpar{\colorbox{orange}{\textcolor{white}{\faInfoCircle}}}
}

\section{Introduction}

Recently a new way to obtain magnetic levitation --- i.e. the hovering of an object in the air due to magnetic forces --- was discovered : a fast rotating permanent magnet (the rotor) being able to `lock in' another magnet (termed here the levitating magnet)~\cite{ucar2021}. This differs from a number of other well-known techniques allowing for magnetic levitation. For example, electromagnetic suspension, where an active control system using an electromagnet stabilizes a permanent magnet, can be used with high speed machinery~\cite{jansen2008magnetically}. Another approach is generating an opposing magnetic field either using eddy currents in conductive materials in the case of electrodynamic suspension (for example in the Maglev train~\cite{maglev}) or directly using diamagnetic materials, or exploiting the Meissner effect of superconductors~\cite{supreeth2022review}. One can also mechanically limit the degrees of the freedom of a levitating magnet, use a rotating magnetic quadrupole to form a magnetic Paul trap~\cite{Pedriat2023} or make stabilizing use of the gyroscopic torque as in the case of the Levitron~\cite{simon1997spin,jones1997simple,michaelis2014horizontal}.

The magnetic levitation by dipole rotation is simple to observe and the experiment is easily doable by students or amateurs but is surprisingly difficult to explain and describe quantitatively. In 2021, H. Ucar published a first general description of this levitation technique~\cite{ucar2021}. His seminal article provided an overview of the phenomenon, including a compendium of experimental realizations, and laid out the main physical ingredients to explain it. The phenomenon started gaining interest after being exposed to the general public~\cite{videoLevitationSansExplications, videoTheActionLab} and notably one of the subjects of the International Physicists'Tournament 2023 consisted in investigating the limitations of the phenomenon~\cite{IPT}. 

A more systematic study was recently published by Hermansen et al~\cite{hermansen2023}. Its authors focus
experimentally on the case of a centered rotor whose magnetic moment
is normal to the rotation axis, observing semi-stable states of finite
lifetime. Notably, they measure the lifetime of the levitation,
describe how it stops and study the influence of the magnet size and
magnetization on the minimum rotation rate for levitation. 
\revise{
The dynamics is further explored by simulations of a model based on the same ingredients as Ucar, which reproduce levitation provided the rotor is slightly shifted off-axis or tilted and dissipation is added at least initially.}
\revisecross{They
propose a mathematical model of the magnet levitation, whose
simulation requires the rotor to be shifted off-axis, and very strong
dissipation to be introduced, in order to recover a levitating regime.
The angular velocity of the levitating magnet around its moment tends
to zero in the simulation while an inspection of their supplementary
movies and our observations show otherwise. Although the simulation qualitatively recovers some observations, it
falls short of clarifying the basic mechanisms at work. More
importantly, the study lacks quantitative comparison between the
advanced model predictions and experimental measurements, for instance
concerning the levitating distance or the polar angle. The gyroscopic
stabilization that is put forward is shown below to be negligible.}

\revisecross{Although both the works of Ucar and of Hermansen et al. agree on the
fundamental physical mechanism behind the levitation, we find that
these studies do not provide a simple general explanation based on
physical principles and experimental observations.}

While both the works of Ucar and of Hermansen {\etal} properly identify the
fundamental physical mechanism behind the levitation, and resort to numerical simulation of the derived complex evolution equation for the translational and rotational degres of freedom of the levitating magnet, neither provides a quantitative comparison of the derived model to experimental measurements.
 
The goal of this
paper is to provide a synthetic physical picture of the phenomenon, \revise{and to back it using quantitative experimental evidence}. We
propose a model based on physical ingredients in line with the
preceding literature, valid for small tilt angles $\gamma$ and
$\theta$ of both magnets with respect to an orthogonal configuration
{(fig.~\ref{fig:schemanip})} but otherwise general, that \revisecross{is more 
straightforward and} yields scaling laws without the need for numerical
simulation. {In particular we characterize the axial equilibrium position.} Confronting the scaling laws and some analytical
predictions to observations, we provide the first quantitative
comparison between experimental data and an analytic model of the
levitation. This allows us to validate the levitation mechanism
suggested by Ucar.

\section{The levitation mechanism}

\subsection{Qualitative description and notations}

\begin{figure*}[!ht]
\centering
\includegraphics[width=\textwidth]{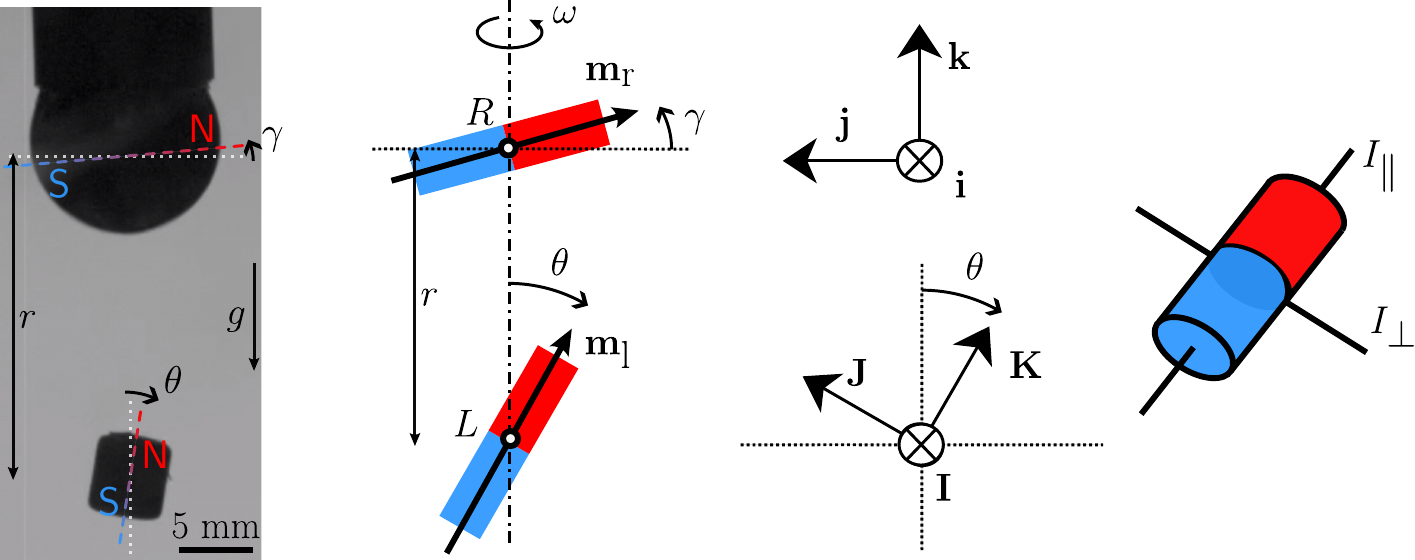}
\caption{Left, Instantaneous picture of the experiment taken with a high speed camera. The top magnet is rotated at rate of $f = \SI{216 \pm 2}{\hertz}$ and its dipolar moment $\mr$ is inclined by an angle $\gamma$ relatively to the plane of rotation (the rotation axis is vertical here). The bottom magnet levitates at a distance $r$ from the top magnet and its dipolar moment $\ml$ makes an angle $\polarAngle$ with the vertical axis. For this experiment, the top magnet is a sphere of diameter \SI{12.7}{\milli\metre} and the levitating magnet is a cylinder of radius \SI{5}{\milli\metre} and height \SI{5}{\milli\metre}. Other magnet shapes (sphere, cylinder, cube ...) both for rotating and levitating magnet can produce the levitation effect, as noted previously~\cite{ucar2021} (see also Supplementary Movies S2, S3 \cite{suppmatt}). Right, schematics and geometrical notations for the model. The basis $(\vb{i}, \vb{j}, \vb{k})$ is attached to the rotating magnet with $\vb{k}$ aligned with the rotation axis and the magnetic moment in the $(\vb{j}, \vb{k})$ plane. The basis $(\vb{I}, \vb{J}, \vb{K})$ is attached to the levitating magnet, with $ \vb{I}=\vb{i}$, and $\vb{K}$ along the magnetic moment. $\Ipar$ and $\Iperp$ are the moments of inertia, parallel to the dipolar moment and perpendicular, respectively.}
\label{fig:schemanip}
\end{figure*}

The typical experimental setup needed to observe the magnetic
levitation in the regime studied here is as follows. One magnet
(termed rotor magnet, or rotor), is fixed and rotates at a rate $\omega$,
typically around \num{150} to \SI{300}{\hertz}, around a vertical axis
to which its magnetic moment is almost (yet not rigorously)
perpendicular. The angle between the horizontal plane and the magnetic
moment of the rotor is noted $\gamma$ and is \revise{a} few degrees. Below the
rotor, another magnet (the levitating magnet, or levitator) hovers
mid-air (video S1 \cite{suppmatt}). It is inclined through an angle $\polarAngle$
compared to the vertical. In this levitating regime, both magnetic
moments remain comprised in the same plane, evidencing the
synchronization between the rotation of the rotor and the precession
of the levitator. We have checked experimentally that this synchronization is respected. A picture of the plane taken from the video, as well as a schematics are displayed in Fig. \ref{fig:schemanip}, showing that \revise{at} any given instant the north pole of the levitator points on the side of the north pole of the \revise{rotor magnet} (i.e.\ $\polarAngle >0 $).

Experimentally, the levitator remains at a fixed distance from the rotor for several tens of seconds when no dissipation is present, to minutes or even hours when in presence of dissipation \revise{obtained by placing an aluminium block nearby the levitating magnet to enhance eddy current damping}. But even the shortest lifetimes, around 10 seconds, are very long compared to one period of rotation (around 5 ms). Thus, the levitation can be considered stable, lasting for thousands of cycles before eventually failing. 
In the present work, we focus on the existence of this stable position.

In this article, we will always consider that the dipolar magnetic moment $\ml$ of the levitating magnet is localized at the center of mass and aligns with a principal axis of inertia (Fig.~\ref{fig:schemanip}). We write $\Ipar$ the moment of inertia around this axis and $\Iperp$ the other two principal moments which we suppose identical. The levitation can be realized for a variety of ratios $\Iperp/\Ipar$, as can be seen in Supplementary Movie S3 \cite{suppmatt}. The model applies whenever the magnetic field of the levitating magnet is dipolar in first approximation, including non spherical magnets such as cylinders or cubes, as can be seen on Supplementary Movie S2 \cite{suppmatt}.

For the convenience of forthcoming calculations, we define two orthonormal bases, both rotating around the vertical (magnet-to-magnet) axis synchronously at rate $\omega$.  The basis $(\vb{i}, \vb{j}, \vb{k})$ is so that $\vb{k}$ is pointing up and the rotor magnetic moment is in the plane spanned by $(\vb{j}, \vb{k})$. The basis $(\vb{I}, \vb{J}, \vb{K})$ is attached to the levitating magnet: $\vb{K}$ points along its magnetic moment and principal inertial axis, inclined by an angle $\polarAngle$ from the vertical, and $\vb{I} = \vb{i}$.

We consider the center of the rotor to be vertically aligned with that of the levitating magnet (i.e. no horizontal offset for either magnets) at a distance $r$. Both magnetic moments turning around the vertical axis at constant rotation rate $\omega=2\pi\, f$ (as discussed below, as opposed to the rotor which is rotating, the levitator precesses with often negligible spin). The rotor is modeled as a perfect magnetic dipole of moment $\mr$ contained in the $(\vb{j}, \vb{k})$ plane and inclined from the horizontal by a small angle $\gamma \ll 1$ (Fig.~\ref{fig:schemanip}). 

\subsection{Counteracting forces}

Let us first offer a qualitative explanation for the existence of a stable point. The reason the levitator stays at a fixed distance to the rotor is because it is constrained there by a repulsive and an attractive force, both of magnetic origin. In general, these magnetic forces dominate over gravity, which is why the levitation can be maintained even on an upside-down configuration, as shown in previous literature~\cite{ucar2021}.

\begin{figure}[!ht]
\centering
\includegraphics[width=\figwidth]{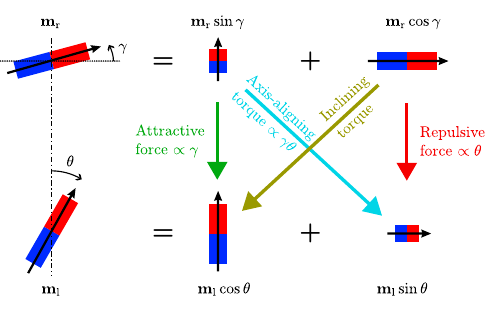}
\caption{Schematics of the magnetic interactions (forces and torques) between the rotor magnet and the levitating magnet.}
\label{fig:magneticinteraction}
\end{figure}

The attractive force comes from the fact that the rotor is slightly inclined, the vertical component $\vb{m_r}\cdot \vb{k}$ generating a vertical magnetic field.
Given the relative orientations of the magnets (see Fig.~\ref{fig:magneticinteraction}) this component of the magnetic interaction is attractive and proportional to $\gamma$.

The repulsive force comes from the slight inclination of the levitator magnetic moment ($\polarAngle$), whose horizontal component interacts with the horizontal magnetic field generated by the rotor. Seeing that the north and south poles of both magnets face each other (see Fig.~\ref{fig:magneticinteraction}), this interaction is repulsive, like the one between two parallel dipoles. Its intensity is proportional to the polar angle $\polarAngle$.

Aside from the qualitative explanation, one can rigorously compute the forces acting between the two dipoles. The magnetic field of the rotor magnet at the center of mass of the levitating magnet reads~\cite{Jackson1999}:
\begin{align}
\label{eq:magfield}
\vb{B_\text{r}}(L) &= \frac{\mu_0 \mr}{4\pi r^3}(\cos \gamma\ \vb{j} + 2\sin\gamma\ \vb{k}) \, .
\end{align}
Thus the magnetic force acting on the levitating magnet is:
\begin{align}
\label{eq:force}
    \vb{F} &= \grad (\vb{\ml}\cdot  \vb{B_\text{r}}(L)) \nonumber \\
    &= \frac{3\csteM}{r^4}
      ((2\sin\gamma\cos\polarAngle-\sin\polarAngle\cos\gamma) \vb{k} + \cos(
      \polarAngle - \gamma ) \vb{j}) \nonumber  \\
    &\approx \frac{3\csteM}{r^4} ((2\gamma-\polarAngle) \vb{k} + \vb{j})
             \qqtext{with} \csteM = \frac{\mu_0 \mr \ml}{4\pi} \, .
\end{align}
The direction $\vb{j}$ rotates at rate $\omega$ in the inertial laboratory
frame, so that the corresponding force component has vanishing
time-average. The corresponding orbiting of the levitating magnet in
the plane normal to the axis is imperceptible experimentally, due to
inertia at these high rotation rates. We therefore neglect this motion
in the following.

The vertical force comprises two components of opposite directions. In
our experiment $\gamma$ is fixed, but we observe experimentally that the levitator tilt $\polarAngle$ strongly varies with the distance $r$ between the two magnets. An equilibrium position for the motion along the vertical axis is reached when
\begin{align}
\label{eq:conditionangle}
    \polarAngle = 2\gamma
\end{align} is verified. 
At this stage however the inclination of the levitator $ \polarAngle$ remains unknown, so that eq.~\eqref{eq:conditionangle} 
tells us little about the equilibrium position for the levitator.
In order to predict the equilibrium distance, one needs to understand the relationship between the polar angle, $\polarAngle$, and the distance, $r$. We will now show that the torque balance provides this dependency.

\subsection{Torque balance}

To study the inclination of the levitating magnet one can work in the frame of reference in rotation at rate $\omega$. The angle $\polarAngle$ is given by the equilibrium between all the different torques acting on the levitating magnet. 
The magnetic torque $\vb{\Gamma}$ can be decomposed into two contributions. The horizontal component of the rotor magnetic moment and the vertical component of the levitator magnetic moment interact through the horizontal magnetic field generated by the rotor, inducing an \emph{inclining} torque, which tends to align the levitator in the reverse horizontal direction, so that the opposite poles of both contribution face each other. 

In contrast, the small vertical magnetic moment of the rotor creates a vertical magnetic field, and its action on the horizontal magnetic moment of the levitator creates an \emph{axis-aligning} torque, that tends to align the levitator with the rotation axis. Since this contribution is proportional to both $\gamma$ and $\polarAngle$, which are chosen to be small, it is always negligible compared to the inclining torque.
\begin{align}
  \label{eq:torque}
    \vb{\Gamma} &= \vb{\ml}\cp \vb{B_\text{r}}(L) \nonumber \\
    &= \frac{-\csteM}{r^3}\ \qty(\cos \gamma \cos \polarAngle + 2 \sin\gamma \sin\polarAngle)\ \vb{i} \qqtext{with} \csteM = \frac{\mu_0 \mr \ml}{4\pi} \nonumber \\
    &\approx \frac{-\csteM}{r^3}\ (1+2\gamma \, \polarAngle)\ \vb{i} \, . 
\end{align}

If this torque were static, the magnet would align with the field on a timescale $\sqrt{\Iperp/\Gamma}$. The observed levitator orientation almost aligned with the rotation axis is possible only if the torque changes orientation on a shorter timescale. The minimum rotation rate should therefore scale as $\sqrt{\Gamma/\Iperp}$. Using the magnetic moment of the magnet used and typical levitation distances gives a minimum rotation frequency of $\sim \SI{100}{\hertz}$, which is the correct order of magnitude.

We now need to apply the laws of mechanics of an axially-symmetric rigid body rotating around a fixed point~\footnote{Which are formally identical to the ones governing the movement of a symmetric spinning top.}. By applying the angular momentum theorem to the levitating magnet at point $L$ (see Fig.~\ref{fig:schemanip}) in the rotating frame of reference \cite{landau1960mechanics}, we obtain
\begin{align}
\label{eq:toupieK}
    \vb{\Gamma}\cdot\vb{K}  &= \Ipar \dv{\spin}{t} \qqtext{with} \spin = \omega \cos \polarAngle  + \omega_K \\
    \label{eq:toupieJ}
    \vb{\Gamma}\cdot\vb{J} &= \Iperp (\dot\omega \sin\polarAngle + 2\omega \dot\polarAngle\cos \polarAngle) - \Ipar \dot\polarAngle \spin \\
    \label{eq:toupieI}
    \vb{\Gamma}\cdot\vb{I} &= \Iperp \ddot \polarAngle + \Ipar \omega \spin \sin\polarAngle - \Iperp \omega^2 \cos\polarAngle \sin\polarAngle \, .
\end{align}
The quantity $\spin$ is the levitating magnet spin, i.e. the angular rotation frequency of the magnet around its own magnetic moment, in the \revise{laboratory} frame of reference. It is accessible experimentally for small polar angle $\polarAngle$ and is always observed to be considerably smaller than the rotor frequency $\omega$. Indeed, levitation can be obtained for a levitator with no or even reverse spin. When unconstrained however, the levitator will eventually start spinning in the same direction as the rotor. Yet the spinning rate of the levitator remains 5 to 10 times smaller than that of the rotor.
Note that, since we are interested in the rapid (on the timescale of one cycle) dynamics of the evolution of $\polarAngle$, we consider an inviscid situation with no dissipation. Air drag, as well as dissipation due to eddy currents, need not be taken into account on such short timescale, where inertia and magnetic actions largely dominate.
The quantity $\omega_K$ represents the speed of rotation of the magnet around its magnetic moment in the rotating frame of reference turning at speed $\omega$, in which the direction of the magnetic moment is fixed.

Due to the fact that the magnetic torques cannot be colinear to the magnetic moment of the levitator, we have
\begin{align}
    \dv{\spin}{t} = 0 \, ,
\end{align}
so that the quantity $\spin$ is conserved. Indeed, we always observe experimentally that $\spin$ is constant (with $\sim \SI{2}{\percent}$ margin) for all the duration of an experiment. We always observe $\spin$ to be inferior to $\omega$, being often negligible. It is also possible to change this quantity using a string attached to the levitator or a static magnet on the side of the experiment, and still observe levitation. This justifies in the following that we neglect the role of $\spin$. But in the general case we have $\spin \neq 0$, as can be seen in the Supplementary Movies we provide \cite{suppmatt} (Supp. Mat. movies S1, S2 and S3) or those from Hermansen et al.~\cite{hermansen2023}, even though their model predicts $\spin$ to be null at all times. 

Multiplying Eq.~\eqref{eq:toupieJ} by $\sin\polarAngle$ gives us:
\begin{align}
    \dv{L_k}{t} = 0 \qqtext{with} L_k = I_\perp \omega\sin^2\polarAngle + \Ipar \spin \cos\polarAngle \, ,
\end{align}
which is a second conserved quantity, corresponding to the \revise{angular momentum} along the axis $\mathbf{k}$. And since both $\omega$ and $\spin$ are unchanging, we necessarily have $\polarAngle = \text{cst}$. In the inviscid situation that we placed ourselves in, valid for the shortest timescales, the problem is mathematically extremely constrained and  the inclination of the levitating magnet cannot change significantly.

We can now use Eq.~\eqref{eq:toupieI} to find the value of $\polarAngle$:
\begin{align}
\label{eq:torquesbalance}
    -\frac{\csteM}{r^3} \ \qty(1 + 2 \gamma \polarAngle) &=  (\Ipar\omega \spin - \Iperp\omega^2 ) \polarAngle
\end{align}
On the left-hand term of Eq.~\eqref{eq:torquesbalance}, we see the magnetic torque formed of the two contributions discussed earlier. On the right-hand term we find the inertial torque, which comes from the fact that we placed ourselves in the rotating frame of reference. There are two contributions to the inertial torque.

\begin{figure}[!ht]
\centering
\includegraphics[width=\figwidth]{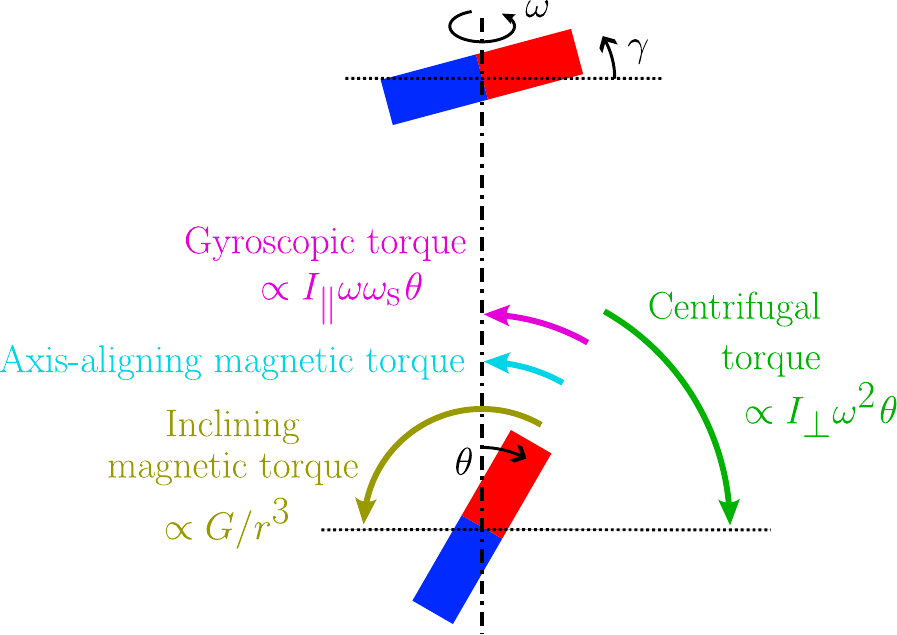}
\caption{The different torques in play. The magnetic torque is the sum
  of an inclining torque, which tends to lay the magnet down to
  $\polarAngle = -\pi/2$, and an axis-aligning torque, which pushes $\polarAngle$ to $0$. Since we are in the rotating frame of reference, we see torques of inertial origin : the gyroscopic torque which, as for a spinning top, straightens up the magnet to the vertical, and the centrifugal torque i.e. the torque resulting from centrifugal forces, that tends to incline the levitating magnet.}
\label{fig:torques}
\end{figure}

The leftmost term consists of the \emph{gyroscopic} torque, proportional to the axial moment of inertia, which tends to incline the levitator towards the vertical axis. This is the term that governs the equilibrium of spinning tops and gyroscopes, but here this term is not dominating \revise{as $\spin$ is usually small.} Indeed at first order the levitating magnet is not properly \emph{spinning} around the vertical axis: rather its magnetic moment is {precessing on a cone} around the vertical axis, but \revise{a point on the equator of }the levitator \revise{hardly spins around said axis}. \revisecross{This is one of the original features of this kind of magnetic levitation, and a fundamental difference compared to the levitron: here the levitator is not a gyroscope and thus contrary to what has been written previously \cite{hermansen2023} the equilibrium of the whole situation is \emph{not} due to gyroscopic effects, which in our case are destabilizing.}

The last term consists of the \emph{centrifugal} contribution of the inertial torque. It is the torque generated by the addition of all the centrifugal forces along the levitator, which is proportional to the transverse moment of inertia and tends to lay down the levitator in the horizontal plane perpendicular to the rotation axis. \revise{Note that this torque is termed `gyroscopic' in ref.~\cite{hermansen2023} because of formal resemblance, in the rest frame of the levitator where it has spin $-\omega$, thus slightly stretching the textbook usage as a
torque linked to proper spin~\cite{Timoshenko_1948,
  fowles_analytical_1997, yehia_2022, perez_mecanique}. We favor a
distinct naming to stress an original feature of this kind of magnetic levitation, and a fundamental difference compared to the levitron: here the levitator is not a gyroscope, and gyroscopic effects, if any, are destabilizing.}

Using $\polarAngle, \gamma \ll 1$ and $\spin \ll \omega$, we can keep only the dominating terms of Eq.~\eqref{eq:torquesbalance} to see that the value of $\polarAngle$ is given by a balance between the centrifugal inertial torque and the inclining magnetic torque as illustrated in Fig.~\ref{fig:torques}. We deduce from this the dependency of $\polarAngle$ in $r$ :
\begin{align}
\label{eq:scalinglaw}
    \polarAngle &\approx \frac{\csteM}{r^3}\frac{1}{\Iperp\omega^2} \, .
\end{align}
At short times, the angle of inclination of the levitating magnet thus varies as $1/r^3$. 

\subsection{Conclusion on the levitation mechanism}

\begin{figure}[!ht]
\centering
\includegraphics[width=\figwidth]{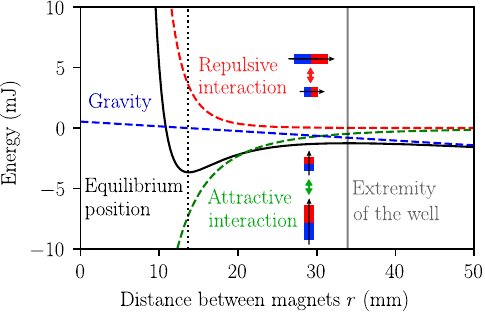}
\caption{Potential energy diagram as a function of the distance between magnets $r$.{This corresponds to the point $\omega = 1362$~rad/s of figure \ref{fig:fit-r-omega}.} The equilibrium position results from an equilibrium between an attractive term $\propto -1/r^3$ and a repulsive term $\propto \polarAngle/r^3$. Since $\polarAngle \propto 1/r^3$, the repulsive interaction potential energy is as $1/r^6$. The equilibrium position is at the minimum of the energy well, where small oscillations can take place. When the levitator is below the rotating magnet, gravity makes the situation metastable and beyond a critical distance the levitator will run off to $r \rightarrow \infty$. Note that if the levitator is placed above the rotor, the metastability turns into a complete stability \revise{for vertical displacements} as gravity acts in the opposite direction. 
}
\label{fig:energydiagram}
\end{figure}

The levitation takes place because of the balance between the attractive component of the magnetic force, which only depends on the constant inclination angle of the rotor $\gamma$, and its repulsive component, which depends on the inclination angle of the levitator $\polarAngle$. According to the torque balance condition, on short timescales we always have $\polarAngle(r) \propto 1/r^3$. Therefore, we can now draw the whole physical picture, using a potential energy diagram depicted Fig.~\ref{fig:energydiagram}.

In this Figure, the repulsive and attractive components are represented and the resulting potential energy is plotted \revise{as} a function of the distance between the two magnets. One can see that the energy landscape exhibits a potential well which defines the stable position for the levitator. For completeness, the effect of gravity was added. 

When the magnets are close, the magnetic torque increases, and to keep the equilibrium the centrifugal torque also augments, so the levitating magnet leans towards the horizontal. This effect strongly increases the repulsive magnetic force. When the levitating magnet is further away, the attractive magnetic force gains importance, until an equilibrium position is attained. The addition of gravity deforms the energy well, and makes the situation metastable, as illustrated in Fig.~\ref{fig:energydiagram}.

\newpage
\section{Experimental verification}

In this section, we present our experimental results and quantitatively discuss their agreement with the model developed above.

\subsection{Torque balance}

Experiments in this subsection were conducted using as a rotor a \SI{12.7}{\milli\metre} diameter spherical magnet of remanence \SI{1.32\pm 0.03}{\tesla} glued to an aluminum bit fitted in the chuck of a motor tool (\textsl{Dremel}) and spun between 12000 and 18000 rpm. Its magnetic moment is inclined by an angle \revise{$\gamma=\SI{6\pm 1}{\degree}$} from the horizontal.
The levitating magnet was a cylinder-shaped magnet (\SI{5}{\milli\metre} height, \SI{5}{\milli\metre} diameter) weighting \revise{\SI{746}{\milli\gram}} with a remanence of \SI{1.35\pm 0.02}{\tesla}. \revise{During the initialization of the levitation we used an aluminium block as induction damper which was removed once the levitation has started.}

The setup was back-lit by a LED panel and a small portion of wire of negligible weight was attached to the levitating magnet to keep track of the rotation around its own axis. In order to resolve all the time scales involved, videos were taken with a high-speed camera, Chronos CR14-1.0 at a frame rate of 8810 frames per second. The images were analyzed using in-house \textsl{Python} routines in order to extract the distance between the magnets $r$ and the polar angle $\polarAngle$.

\begin{figure}[!ht]
\centering
\includegraphics[width=0.95\figwidth]{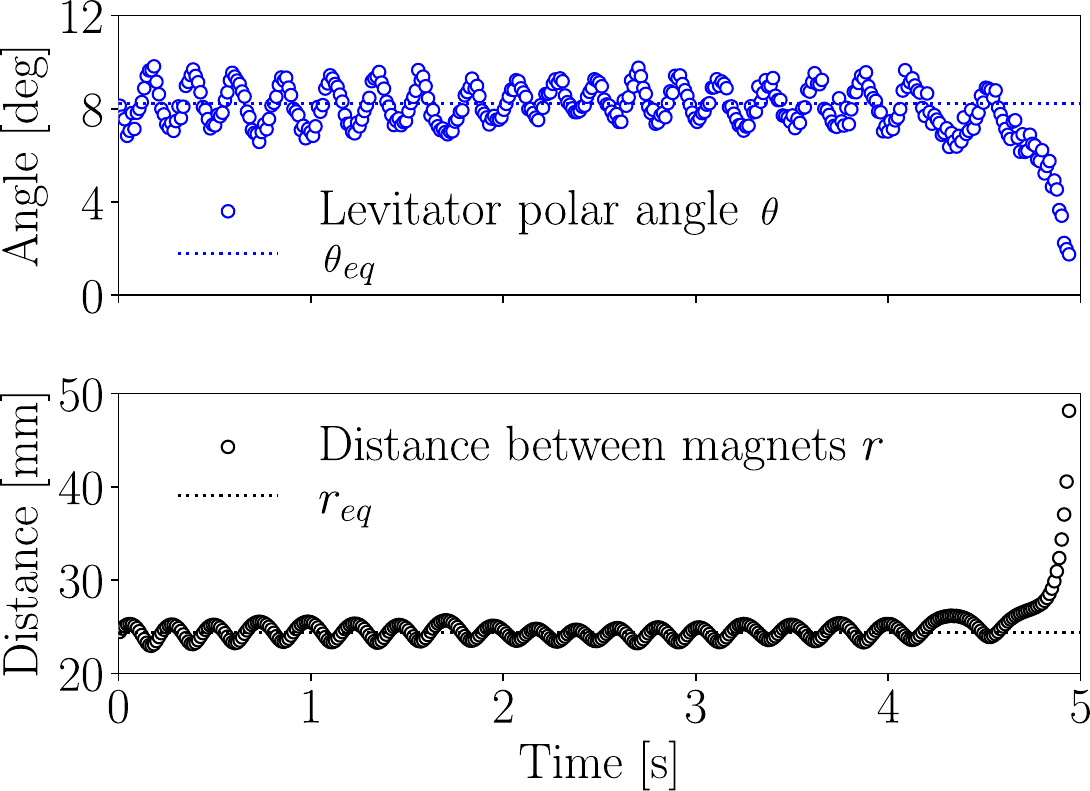}
\caption{Angle of the levitating magnet and distance between the two magnets as a function of time for one given rotor frequency. For this recording the rotor was turning at $f = \SI{216 \pm 2}{\hertz}$.}
\label{fig:expdata}
\end{figure}

An example of the data obtained can be seen on figure
\ref{fig:expdata}. The levitating magnet is in a bound state for more
than 4~s, as it performs small oscillations around the equilibrium
position\revise{ (in a potential well similar to that of Fig.~\ref{fig:energydiagram}, a representation of which is given in the Supplementary Material S4 \cite{suppmatt})}. One can easily observe that the levitating distance $r$ and the polar angle $\polarAngle$ are highly correlated. Near the end of the recording, we see the magnet falling vertically as $r$ increases towards infinity. 

The data allows us to validate the scaling law of Eq.~\eqref{eq:scalinglaw}, arising from the equilibrium of magnetic and inertial centrifugal torques in a conservative system, along with approximations that are appropriate in our set-up. The period of the oscillations (typically 0.2 s) is much greater that than of the imposed rotation (smaller than \SI{5}{\milli\second}), leaving enough time for the inclination of the levitator, $\polarAngle(t)$ to adapt to a varying distance $r(t)$.

\begin{figure}[!ht]
\centering
\includegraphics[width=0.95\figwidth]{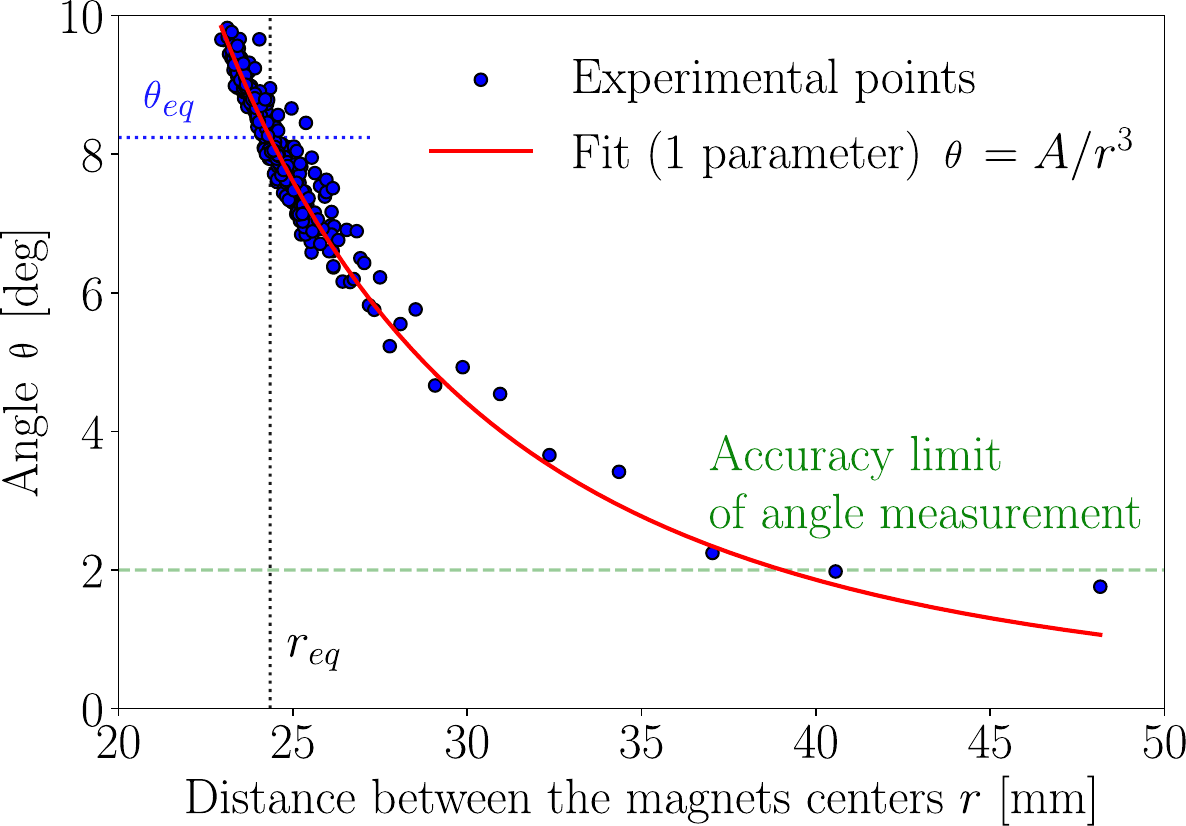}
\caption{The inclination angle of the levitation magnet as a function of the distance between the two magnets,  for a rotor frequency of $f = \SI{216 \pm 2}{\hertz}$.  According to our theoretical model (Eq.~\eqref{eq:scalinglaw}), we expect a power law of exponent -3. The experimental values are well fitted by $A/r^3$ red solid line, where $A$ is an adjustable parameter \revise{(see text)}. The equilibrium values for the distance $r_{eq}$ and the inclination $\polarAngle_{eq}$ are indicated in dotted lines. By performing small oscillations, the levitating magnet explores the potential well, while verifying the scaling law. During the fall of the magnet, the scaling law is still respected.}
\label{fig:fitr3}
\end{figure}

The data in figure~\ref{fig:fitr3}, in which the angle $\polarAngle$ is plotted against the distance $r$ between the magnets, is in excellent agreement with the predicted power-law, even during the fall of the levitating magnet. \revise{The fitting parameter takes the value $A_\text{exp} = (2.08 \pm 0.34)\times10^{-6}$~rad.m$^3$, while the predicted value using Eq.~\eqref{eq:scalinglaw} is $A_\text{th} = (2.32 \pm 0.08)\times10^{-6}$~rad.m$^3$, which is compatible with the experiments.} We thus correctly identified the mechanisms behind the scaling law and demonstrates the robustness of our non-dissipative approach for short times.

\subsection{Equilibrium state: polar angle and distance}

Experiments in this subsection were performed using a setup similar to the one presented in the previous subsection. The rotor was a \SI{10}{\milli\metre} diameter spherical magnet with remanence \revise{\SIrange{1.22}{1.26}{\tesla}}. The tilt angle of the rotor was chosen to be $\gamma =
\SI{9 \pm 1}{\degree}$ and the rotation speed was varied. The resulting equilibrium distance and the equilibrium inclination of the levitator were measured using a large block of aluminum which serves as a inductive damper for the oscillations reported in Fig.~\ref{fig:expdata}. \revise{Each data point corresponds to an individual run which was filmed using a high-speed camera. We used \textsl{Python} image analysis tools to extract the distance between magnets as well as the polar angle $\polarAngle$, and an exploitation of the spectrogram of the sound made by the motor tool to determine the rotation frequency.} 

\begin{figure}
    \centering
    \includegraphics{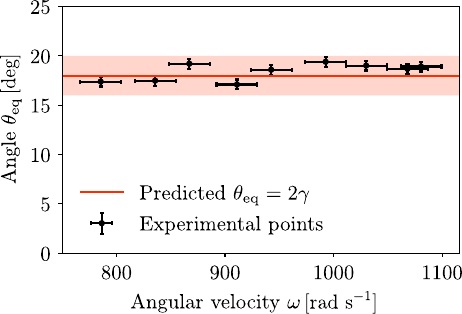}
    \caption{Measured value of the angle $\polarAngle_\mathrm{eq}$ at
      equilibrium for different rotor angular velocities. The angle
      $\gamma$ is such that $\gamma = \SI{9 \pm 1}{\degree}$. We observe that Eq.~\eqref{eq:conditionangle} holds up to experimental precision.}
    \label{fig:phi-2-gamma}
\end{figure}

As a reminder, according to Eq.~\eqref{eq:conditionangle}, the mean polar angle $\polarAngle$ should be independent of the frequency and is equal to twice the rotor inclination $\gamma$. \revise{To confirm this, we used as levitating magnet} a cylindrical magnet with height \SI{12.5}{\milli\metre} and diameter \SI{4}{\milli\metre} of remanence \SIrange{1.29}{1.32}{\tesla}. \revise{Exploiting a video of the experiment, we measured the angle of inclination of the levitator $\theta$. We were limited in precision when measuring the angle because it exhibits small variations due to vertical oscillations.} Figure~\ref{fig:phi-2-gamma} indeed shows that the inclination of the \revise{levitating magnet} does not depend on the rotation speed, and remains equal to $2 \gamma$ with experimental uncertainties. Again, these result confirm the prediction of our dipolar model. 

Combining Eqs.~\eqref{eq:conditionangle} and \ref{eq:scalinglaw}, one finds that the equilibrium condition (in the no gravity limit) reads:  
\begin{equation}
    r=\sqrt[3]{ \frac{\csteM}{2\,\gamma\, I_\perp\, \omega^2}} \propto \omega^{-2/3}\label{eq:equilibiriumposition}.
\end{equation}

\begin{figure}
    \centering
    \includegraphics{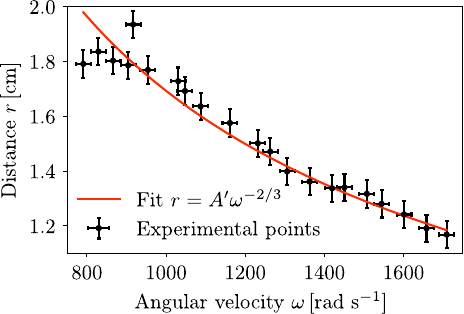}
    \caption{Equilibrium distance $r$ between the two magnets as a function of the rotation speed $\omega$. The experimental data is correctly described by a model $r = A' \omega^{-2/3}$, with $A'$ a free parameter, which validates the theoretical result in Eq.~\eqref{eq:equilibiriumposition} \revise{(see text)}.}
    \label{fig:fit-r-omega}
\end{figure}

The experimental measurements of the equilibrium distance as a function of the rotation speed are plotted in Fig.~\ref{fig:fit-r-omega}. \revise{For these experiments, the levitating magnet was a sphere of diameter \SI{10}{\milli\metre} and remanence \SIrange{1.22}{1.26}{\tesla}. Errors in $\omega$ correspond to the measurement of the rotation speed from video acquisition and errors in $r$ are determined by the fluctuations in position during movement for one acquisition.}
Again, the predicted $\omega^{-2/3}$ power-law is in excellent agreement with the data. \revise{The fitting parameter takes the value $A'_\mathrm{exp} = \SI{1.69(0.02)}{\metre\second\tothe{2/3}}$, while the predicted value using Eq.~\eqref{eq:equilibiriumposition} is $A'_\mathrm{th} = \SI{1.48(0.06)}{\metre\second\tothe{2/3}}$. The disagreement between these two values can be explained by the simplicity of our model, which induces systematic errors. In particular, 
as the magnetic field of the magnets varies strongly at distances less than few radii, the assimilation of magnets to point dipoles is a considerable approximation. Consequently,
the actions felt by the levitating magnet are not exactly the force and torque given in Eqs.~\eqref{eq:force} and~\eqref{eq:torque}. Nevertheless, these two values being close to each other and of the same order of magnitude confirms that the model correctly encompasses the main physical effects.}
Note that levitation can be obtained for lower values of the rotation speed but as $\omega$ decreases, the role played by gravity become increasingly preeminent.

\section{Conclusion / perspective}

In this article, we have proposed a clear explanation of the levitation phenomenon first described by Ucar~\cite{ucar2021}. Our approach focuses on the case in which rotor is slightly tilted from the horizontal, which creates a two-component magnetic interaction, leading to the existence of a stable point. 
Our model is compatible with the approaches of Ucar~\cite{ucar2021} and Hermansen et al.~\cite{hermansen2023}, but instead of relying on numerical simulation of the dynamical system, we used appropriate simplifications and analytic calculations to draw a comprehensible picture of how the levitation emerges from the interactions. Moreover, we derived simple scaling laws that match our experimental data quantitatively.

The theory and results presented here are obtained with working assumptions whose range of applicability needs to be discussed and tested.

First, the scaling given by Eq.~\eqref{eq:scalinglaw} is derived in the
absence of any source of dissipation. This is a reasonable assumption
which holds for low rotation speeds and which remains valid at
short timescales. While it allows for the explanation of the
levitation, arguably in a more realistic approach, encompassing wider
timescales and studying the destabilization mechanism and the lifetime
of the phenomenon, dissipation, coming \revise{either from} air drag or eddy
currents heat loss, should be included \revise{as well as energy injection from the rotor magnet}.

In the regime presented here, as far as we can tell, the two magnetic
moments remain in the same (rotating) plane, one magnet rotating, the
other precessing. In general a lag between the two can exist. Indeed,
a phase angle between the rotor and the levitator of
\SI{6.4\pm5.1}{\degree} was observed by Hermansen \textsl{et
  al}~\cite{hermansen2023}. Such a small phase shift has no
significant impact on our model and its conclusions, as it induces
only a quadratic correction on the magnetic torque
(appendix~\ref{sec:lag}). As a lag induces axial torque (although
quadratically small), it may however be an important degree of freedom
for the dynamics on longer time-scales and for the stability of the
system.

Our model treats the two magnets as point-like dipoles. While it is known that spheres of uniform magnetic material indeed create a dipolar field, the deviation from such a field for non-spherical, or unevenly magnetized levitators may come into play when the magnets get close to one another. Note however that the levitation is possible even when considering strongly non-spherical, and thus less dipolar, levitating magnets (Supplementary Movie S2 \cite{suppmatt}). The limitations mentioned above imply corrections merely of higher order, since our model quantitatively describes well the data.

\revise{In conclusion, we provided a first quantitative experimental verification of the stabilizing mechanism for an original magnetic levitation in a rotating dipolar field, that differs fundamentally from the spin-stabilization of the levitron. Compared to the latter which balances weight with magnetic repulsion, trapping occurs only through magnetic interactions which have both an attractive and a repulsive contribution. The trapping energy in the milli-Joule range (fig.~\ref{fig:energydiagram} and \cite{ucar2021}) is easily two orders of magnitude above that of the levitron, making this levitation far easier to reproduce. Last but not least, here dissipation helps in reaching the equilibrium position, which is maintained as long as the second magnet is spun, while dissipation limits the levitron's duration and stability. }
There are several interesting ways to deepen the understanding of the
phenomenon that we have not explored. We believe that the most
promising one would be to add one (or several) degree of freedom to
the small-angle model to study how the destabilization occurs and what dynamical
path it takes to escape the potential well. It would also be
interesting to have an estimation of the lower bound of the rotor
speed allowing for levitation, which could be quantitatively compared
to experimental measurements such as the ones made by Hermansen et
al.~\cite{hermansen2023}.

For the purpose of open access, the author has applied a Creative Commons Attribution (CC BY) license to any Author Accepted Manuscript version arising from this submission.

\appendix
\section{Effect of levitator precession phase lag}
\label{sec:lag}

Consistent with our experimental observations, we assume that the levitator's magnetic moment is contained in the same vertical plane as the magnetic moment of the rotor. Here we discuss the implications of a small but non-zero constant phase shift between the horizontal components that rotate at the same rate around the vertical. When the magnetic moment of the rotor is aligned with $-\vb{j}$, note $\vb{\ml} = \ml (-\sin\theta \sin\delta \,\vb{i} - \sin\theta \cos\delta \,\vb{j} + \cos\theta \,\vb{k})$ the moment of the levitator, whose projection in the $(\vb{i},\vb{j})$-plane
  lags by an angle $\delta$ behind that of the rotor. With eq.~\eqref{eq:magfield} the magnetic torque then reads
\begin{align}
     \vb{\Gamma} &= \vb{\ml}\cp \vb{B_\text{r}}(L) \nonumber \\
     &= \frac{-\csteM}{r^3}\ ((\cos \gamma \cos \polarAngle + 2 \sin\gamma \sin\polarAngle \cos\delta)\ \vb{i} \nonumber \\
     &\phantom{\frac{-\csteM}{r^3}\ (}- 2 \sin\gamma \sin\polarAngle\sin\delta\ \vb{j} + \sin\gamma \sin\polarAngle\cos\delta\ \vb{k})
     \nonumber \\
    &\approx \frac{-\csteM}{r^3}\ ((1+2\gamma \, \polarAngle)\ \vb{i} 
    + \polarAngle\delta \vb{k})) \qqtext{with} \csteM = \frac{\mu_0 \mr \ml}{4\pi}\,. 
\end{align}
Compared to the torque \eqref{eq:torque} the inclining magnetic torque
$\propto\vb{i}$ is unchanged to second order in small angles, justifying
that we neglect the phase $\delta$. Interestingly there is a second order
term $\propto\vb{k}$ that tends to change $\delta$, which may play a role in
stability and in determining the levitator spin around its axis.

\revise{\section{Levitator orbiting around rotation axis}}
\label{sec:side}

\revise{%
If we consider that the levitator is at $\vb{r} = -z\vb{k} + \di \vi + \dj \vj$ then we can compute the force acting on the levitator. Considering a small deviation from the equilibrium presented above, we take $\gamma, \polarAngle \ll 1$ and also consider $\di, \dj \ll r$. Last, we place ourselves at the equilibrium position on the vertical axis, i.e. we consider eq \eqref{eq:conditionangle} to be valid. We then obtain, up to second order in $\gamma, \polarAngle, \di/r, \dj/r$ : 
\begin{align}
    \vb{F} = \frac{3\,G}{r^4} 
    \bigg(&  
    4\frac{\dj}{r}\,\vk - (2\gamma \frac{\di}{r}+5\frac{\di\,\dj}{r^2})\,\vi \nonumber \\
    &+ (1 + 2\gamma\frac{\dj}{r} - \frac{{\di}^2}{2\,r^2} - \frac{11\,{\dj}^2}{2\,r^2})\,\vj
    \bigg)
\end{align}
In the lateral direction, the leading term is $\vb{F}^{(0)} = 3G/r^4 \,\vj$ which is seen on eq.~\eqref{eq:force}, all other terms being of order two. 
The discussion of the stability of a very small displacement in the lateral direction would thus necessitate a quadratic development of all the equations previously computed and an analysis over long time scales for which dissipation as well as energy injection can become relevant.
{ Note that since it only stays in the same direction during a short time (of order $1/\omega$), the transverse magnetic force $F \approx 3G/r^4\,\vb{j}$ can only move the magnet of a distance of order 
$$
\frac{\dj}{r} \sim \frac{F}{r\,m\,\omega^2} = \frac{6\,\gamma\, \Iperp}{m\, r^2} \approx 6\gamma\qty(\frac{a}{r})^2 \ll 1  
$$ 
where $a$ is the characteristic size of the levitating magnet. Since this displacement is small the magnet stays, to a good approximation, on the rotation axis. As the levitator moves in the opposite direction half a period later, this magnetic force effectively averages at 0.}
}

\end{document}